\documentclass[12pt,preprint,epsfig]{aastex}

\begin{document}
\title{Extended black hole cosmologies in de Sitter space}
\author{Maurice H.P.M. van Putten}
\affil{Le Studium IAS, 45071 Orl\'eans Cedex 2, Universit\'e d'Orl\'eans, France}

\begin{abstract}
We generalize the superposition principle for time-symmetric initial data of black hole
spacetimes to (anti-)de Sitter cosmologies in terms of an eigenvalue problem
$\Delta_g\phi=\frac{1}{8}(R_g-2\Lambda)\phi$ for a conformal scale $\phi$ applied to a
metric $g_{ij}$ with constant three-curvature $R_g$. Here, $R_g=0,2$ in the 
Brill-Lindquist and, respectively, Misner construction of multihole solutions for $\Lambda=0$. 
For de Sitter and anti-de Sitter cosmologies, we express the result for $R_g=0$ in 
incomplete elliptic functions. The topology of a black hole in de Sitter space can be extended 
into an infinite tower of universes, across the turning points at the black hole and 
cosmological event horizons. Superposition introduces binary black holes for small separations 
and binary universes for separations large relative to the cosmological event horizon. The 
evolution of the metric can be described by a hyperbolic system of equations with curvature-driven 
lapse function, of alternating sign at successive cosmologies. The computational problem of 
interacting black hole-universes is conceivably of interest to early cosmology when 
$\Lambda$ was large and black holes were of mass $<\frac{1}{3}\Lambda^{-1/2}$, here facilitated
by a metric which is singularity-free and smooth everywhere on real coordinate space.
\end{abstract}

\section{Introduction}

The multihole solutions of Brill-Lindquist and Misner \citep{mis63,bril64,coo01}
arise out of a superposition principle in the Hamiltonian constraint of the Einstein
equations for time-symmetric initial data. Here, the Schwarzschild
line-element in coordinates $(r,\theta,\varphi)$ is transformed into an
isotropic line-element described by a conformal factor $\phi(r)$, giving
rise to a conformally flat spacetime in vacuum. This construction is
remarkable, in allowing for multihole solutions with different extended 
topologies. To a black hole binary, the Brill-Lindquist solution 
attributes a three-sheet topology, whereas the Misner solution attributes a 
two-sheet topology to the same. 

The embedding of black holes in (anti-)de Sitter space is well-known
for a single black hole with one-sheet topology. De Sitter space is
of interest in view of considerable observational evidence for a small but 
distinctly positive cosmological constant in the $\Lambda$CDM model 
\citep{deb00,han00} with an expected improved uncertainty by the recently
launched {\em Planck} satellite \citep{esa08}. The stability and thermodynamics 
of the cosmological event horizon in the black hole-de Sitter space is  
relevant to its potential role to early cosmology \citep{dav87,cha97}. 

Here, we consider the problem of extending space beyond the cosmological
event horizon. We approach this problem in terms of the turning points 
at extrema of the circumference in the one black hole-de Sitter space.
The result opens the possibility for novel topologies on scales that
reach beyond the visible universe. Space of finite volume further
facilites representation in a metric that is smooth everywhere. This 
is of direct interest to introducing spectral methods for
calculating wave-templates for gravitational-wave observatories LIGO 
and Virgo as they are gradually improving their sensitivity.

To begin, we recal the Brill-Lindquist line-element of a single
Schwarzschild black hole \citep{bril64,coo01}. The event horizon
of a black hole of mass $E$ represents a turning point in the embedding 
in two asymptotically flat sheets, 
\begin{eqnarray}
ds^2 = -\tanh^2(\lambda/2) dt^2 
       + 4E^2\cosh^4(\lambda/2)ds_D^2,
\label{EQN_BM}
\end{eqnarray}
over the donut $ds_D^2=d\lambda^2 + \frac{dx^2}{1-x^2}+(1-x^2)d\varphi^2$
with $-\infty < \lambda < \infty$.
Here, we transformed the Brill-Lindquist line-element with conformal factor
$1+\frac{E}{2\rho}$ in spherical coordinates $(\rho,\theta,\varphi)$, 
$x=\cos\theta$ by $\rho=\frac{1}{2}Ee^{\lambda}$. The M\"obius invariance 
$\lambda\leftrightarrow-\lambda$ comes with opposite signs in the lapse 
fuction $N=\tanh(\lambda/2)$ on either sheet, wherein the horizon surface 
corresponds to the simple zero $N=0$. 

By Liouville's theorem, spacetime singularities are inherent to any black hole spacetime 
which is asymptotically regular. They can be moved away into the complex plane as (\ref{EQN_BM}) 
illustrates by mapping $\rho=-\frac{1}{2}E$ to $\lambda=\pi i$ (mod $2\pi$), whereby the metric is analytic
everywhere for all real and finite $\lambda$, wherein $\lambda=\pm \infty$ represent 
coordinate singularities associated with asymptotic infinity on each sheet. 

The black hole singularity is not directly accessible by observation, whether in
real or complex coordinate space. Even an observer in free fall onto a black hole 
never reaches the central singularity when considered in real coordinates, as in 
the Schwarzschild line element. Upon approaching the event horizon, time-at-infinity, 
$t$, becomes arbitrarily large, which signals evaporation of the black hole by Hawking 
radiation. The observer hereby traces a shrinking event horizon, never to penetrate it, 
during which time the black hole singularity diminishes in strength. Note that this result
only uses evaporation of a black hole in a finite time-at-infinity, wherein the
velocity of the black hole horizon is representative for the luminosity regardless 
of the details of Hawking radiation. Hawking radiation hereby introduces 
invariance to the cosmic censorship conjecture, by treating singularities in real and
complex coordinate space on equal footing. The same arguments shows 
that (\ref{EQN_BM}) defines a non-traversable wormhole.

In de Sitter space, coordinate singularities associated with the infinite extend of 
space are avoided. The result for multihole solutions in de Sitter space, then, is a 
metric which is {\em everywhere} nonsingular and finite on the coordinate cover of
the visible universe with no need for compactification. The cosmological event horizon 
represents an additional length scale, which can be of interest in considering black 
hole binaries with large separations. It introduces an additional turning point which 
opens a window for novel large scale extensions beyond.

In \S2, we derive the formalism for constructing multihole initial data in (anti-)de 
Sitter space. In \S3 we give some illustrative examples for binary black holes and
binary cosmologies. In \S4, we propose a hyperbolic system of equations for their
evolution based on the 3+1 Hamiltonian equations of motion \citep{arn62}, where
hyperbolicity generally facilitates stable numerical implementation (e.g., 
\cite{van96,nag04,cal06} and references therein) by ensuring a real dispersion relation
and hence stability whenever the Courant-Friedrichs-Lewy condition \citep{cou67} is
satisfied. An outlook is included in \S5.

\section{A superposition principle in (anti-)de Sitter space}

\begin{figure}
\centerline{\includegraphics{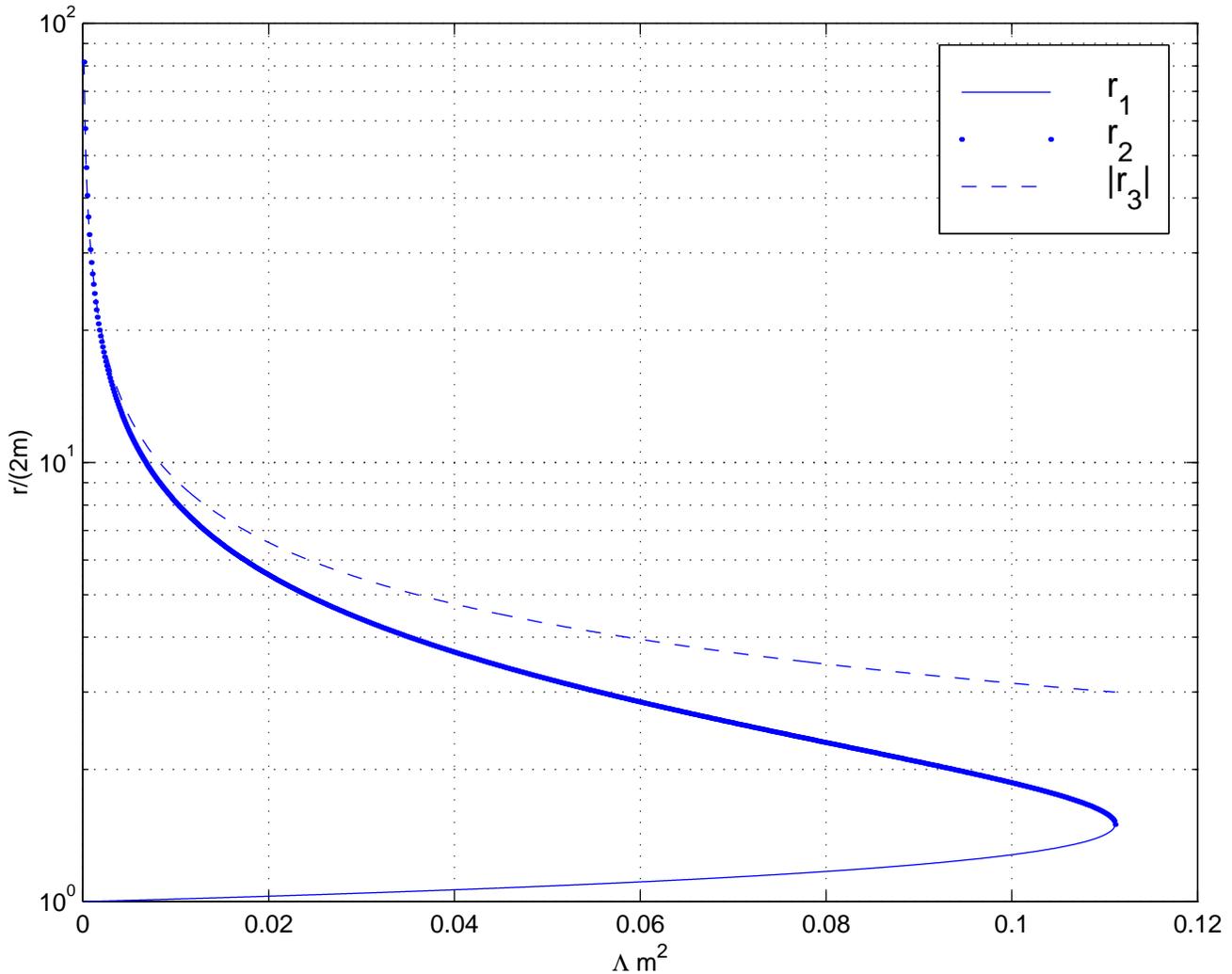}}
\caption{{\small
Shown are the three real roots of the redshift factor in the Schwarzschild-de Sitter space
as a function of $\Lambda m^2>0$. They define the coordinate location of the horizon of the
black hole $(r_1)$ and of the cosmological event horizon $(r_2)$. The roots bifurcate into
a pair of real roots at $\Lambda m^2=1/9$. The maximum mass of a black hole in a de Sitter
space is hereby $1/3\Lambda^{-1/2}$ \citep{pod99}. Shown is further the numerical root
$r_3<0$.}}
\label{EQN_roots}
\end{figure}

The line-element of a Schwarzschild black hole of mass $m$ in a de Sitter space with
cosmological constant $\Lambda$ is
\begin{eqnarray}
ds^2=-(1-2m/r-\frac{1}{3}\Lambda r^2) dt^2 + \frac{dr^2}{1-2m/r - \frac{1}{3}\Lambda r^2} 
     + r^2d\theta^2 + r^2\sin^2\theta d\varphi^2.
\label{EQN_DS}
\end{eqnarray}
When $\Lambda m^2<\frac{1}{9}$, a black hole of mass $m$ can exist within the larger cosmological 
event horizon defined by $\Lambda$ with a black hole horizon 
at coordinate radius $r=r_1$ and an event horizon at $r=r_2>r_1$ corresponding to the
roots of the redshift factor $N=0$, $N^2=1-2m/r-\frac{1}{3}\Lambda r^2$, in (\ref{EQN_DS}). The
roots are shown in Fig. (\ref{EQN_roots}) with the bifurfaction point at $\Lambda m^2=\frac{1}{9}$,
including the non-physical real root $r_3<0$.
The line-element (\ref{EQN_DS}) can be transformed to an isotropic line-element
\begin{eqnarray}
ds^2=-N^2dt^2 + \phi^4 (d\rho^2 + \rho^2d\theta^2+\rho^2\sin^2\theta d\varphi^2)
\end{eqnarray}
according to the conditions
\begin{eqnarray}
\pm \frac{dr}{\sqrt{1-2m/r-\frac{1}{3}\Lambda r^2}} = \phi^2 d\rho,~~r=\phi^2\rho.
\label{EQN_E1}
\end{eqnarray}
For $\Lambda>0$ the three real roots $r_3<0<r_1<r_2$ of $N$ satisfy $r_1=2m(1+\epsilon)$, 
$r_{2,3}=\pm \sqrt{\frac{3}{\Lambda}}(1\mp\delta),$
where $\epsilon=\frac{4}{3}\Lambda m^2+O(\epsilon^2)$ denotes the horizon
surface of the black hole and $\delta = m\sqrt{\frac{\Lambda}{3}}+O(\delta^2)$. 

The solution to (\ref{EQN_E1}) can be expressed following a logarithmic transformation 
$d\lambda=d\rho/\rho$. For $\Lambda>0$, the M\"obius transformation $t=2m/r$ (with no 
reference to time intended) gives
\begin{eqnarray}
\lambda=\left(\int_{t_2}^{t_1}-\int_{t_2}^{t}\right)\frac{d\tau}{\sqrt{\tau^2-\tau^3-4\Lambda m^2/3}}
=2\frac{F(\pi/2,m_1)-F(\phi,m_1)}{\sqrt{t_1-t_3}}
\label{EQN_L1}
\end{eqnarray}
in terms of the three roots $t_i=2m/r_i$ ($t_3<0<t_2<t_1$), $m_1=\frac{t_1-t_2}{t_1-t_3}$,
$\sin\phi=\sqrt{\frac{(t_1-t_3)(t-t_2)}{(t_1-t_2)(t-t_3)}}$, where $F(\phi,m_1)$ denotes 
the incomplete elliptic function of the first kind \citep{abr68}. For $\Lambda<0$, the only 
real root is $r_2$. We proceed with $r=\frac{6m}{1-12 v}$,
\begin{eqnarray}
\lambda=\int_{v_2}^v \frac{dP}{\sqrt{4P^3-P/12+1/216+\Lambda m^2/12}},
\label{EQN_L2}
\end{eqnarray}
giving rise to the Weierstrass elliptic function
$v(\lambda) = P(\lambda, 1/12, -1/216 - \Lambda m^2/12)$ \citep{abr68}. 

\begin{figure}
\centerline{\includegraphics{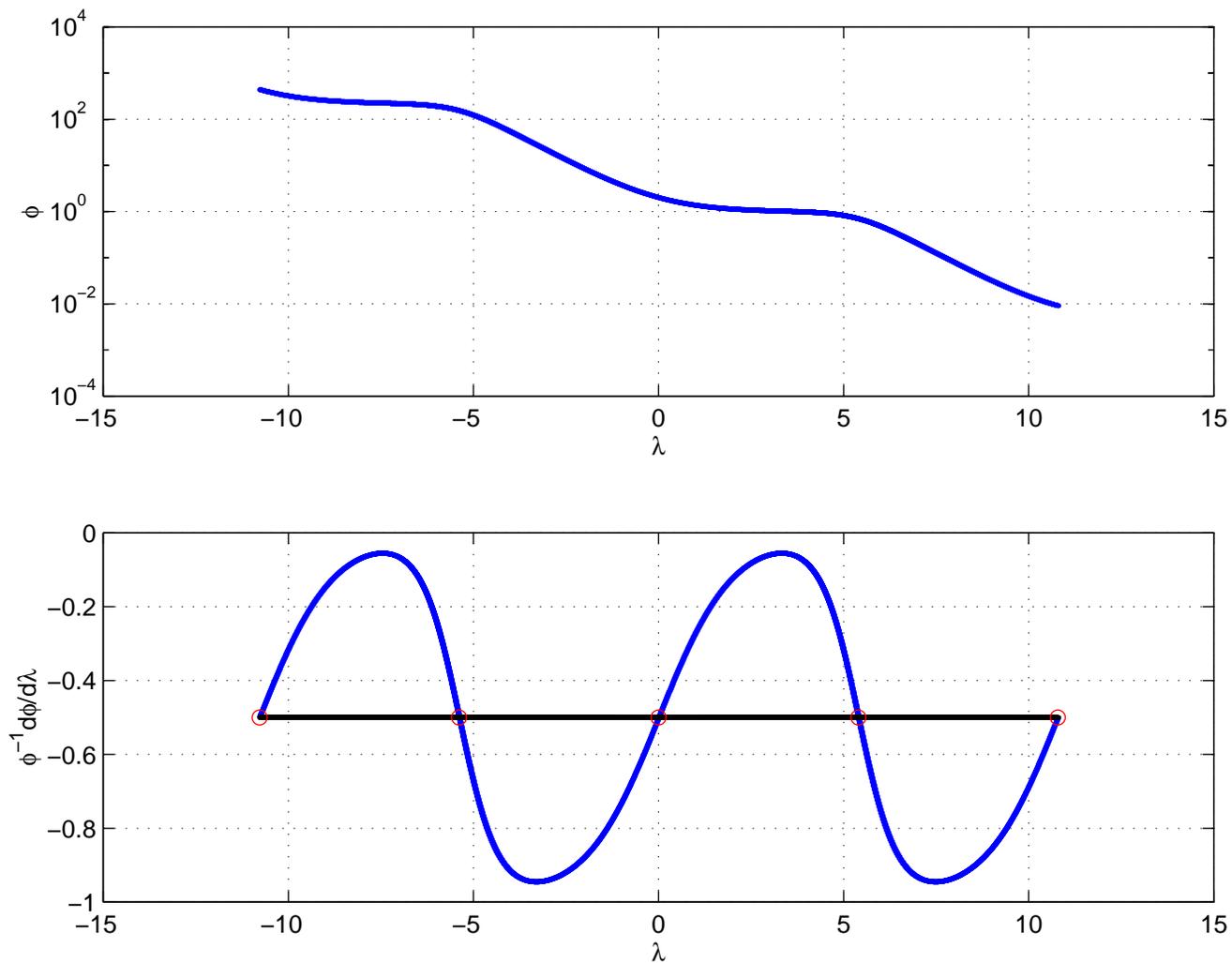}}
\caption{{\small
For a single black hole in de Sitter space, the scale $\phi$ is periodic in
the radial coordinate $-\infty<\lambda<\infty$, representing a ``tower" of cosmologies by 
successive continuations at the turning points defined by the black hole and 
cosmological event horizons. The results shown are computed for $m=1$ and $\Lambda=0.001$,
and may be contrasted with $\phi=\left(1-e^{-\lambda}\right)^{1/2}$ in singular limit $\Lambda=0$.}}
\label{FIG_2a}
\end{figure}

\begin{figure}
\centerline{\includegraphics{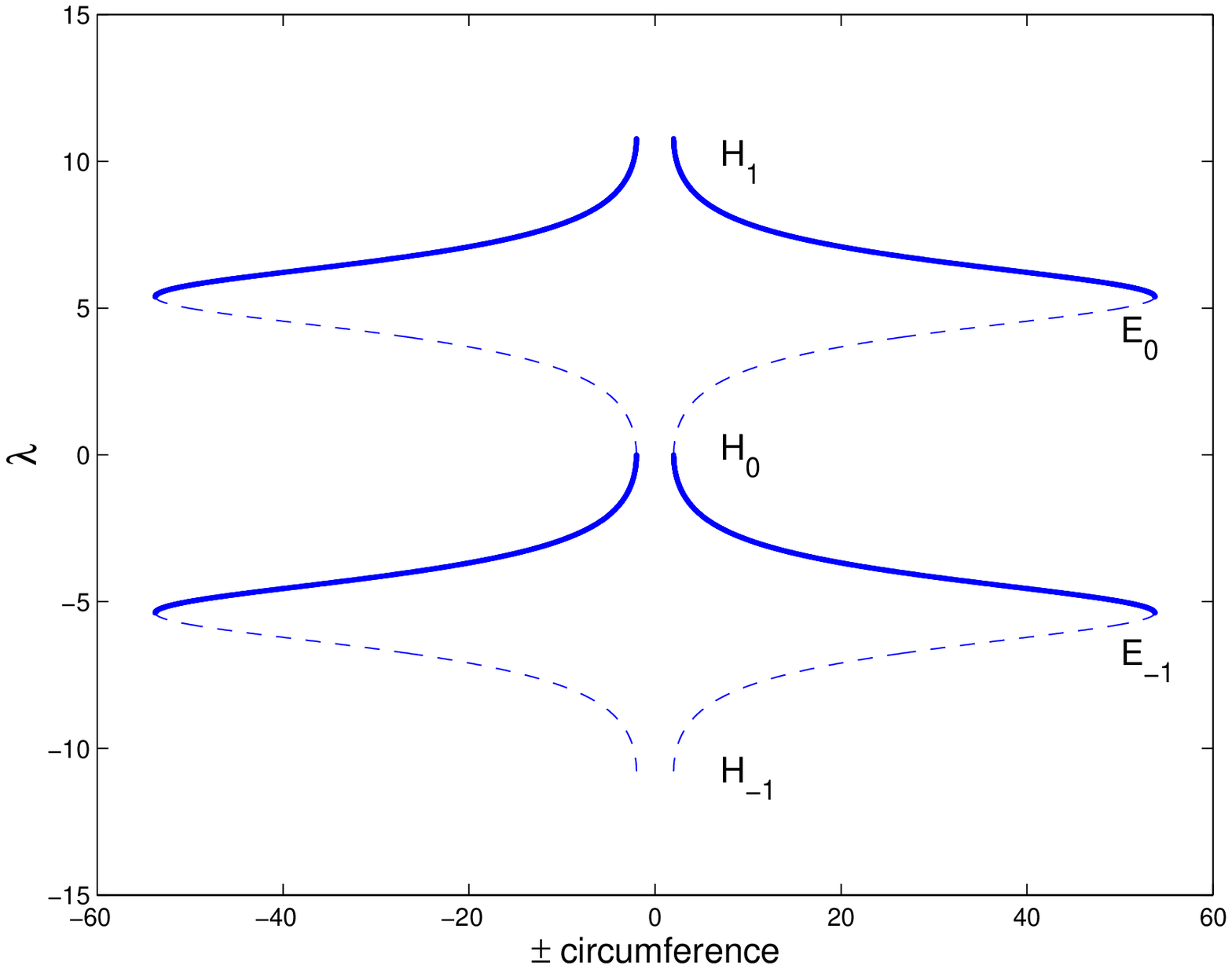}}
\caption{{\small
The extension of black hole-de Sitter cosmology of a single black hole-de Sitter 
spacetime can be realized by joining the eigenfunction solutions at successive 
turning points (solid-to-dashed transitions), representing black hole and cosmological 
event horizons $(H_i, E_i)$. The results shown are computed for $m=1$ and $\Lambda=0.001$.}}
\label{FIG_2b}
\end{figure}

In the resulting conformally flat approach, the resulting extension of de Sitter
black hole spacetime is shown in Figs. (\ref{FIG_2a}-\ref{FIG_2b}). We note that 
the singular limit $\Lambda=0$ reduces to the familiar expressions (e.g. \cite{abr96})
\begin{eqnarray}
\lambda=\ln\left(\frac{1}{m}[r-m+\sqrt{r^2-2mr}]\right),~~r=\rho\left(1+\frac{m}{2\rho}\right)^2,~~
\rho=\frac{1}{4}\left(\sqrt{r}+\sqrt{r-2m}\right)^2.
\end{eqnarray}


We interpret the result as follows.

{\bf Theorem.} {\em The superposition principle for time-symmetric data in the 
Hamiltonian energy constraint generalizes to (anti-)de Sitter space for 
eigenfunctions of the Laplace operator of a scaled metric with constant
3-curvature.}

{\em Proof.} We recall the conformal decomposition of the Ricci tensor for 
$h_{ij}=\phi^{4}g_{ij}$, whereby
$\mbox{}^{(3)}R_{ij}(h)=\mbox{}^{(3)}{R}_{ij}(g) +C_{ij}$
with $C_{ij}=- 2\phi^{-1}[{D}_i{D}_j\phi + {g}_{ij} {\Delta} \phi]
      + 2\phi^{-2}[3 {D}_i\phi{D}_j\phi - {g}_{ij} {D}^p\phi {D}_p\phi].$
Following a reduction of the Ricci tensor from four to three dimensions,
the Einstein equations for time-symmetric data give
\begin{eqnarray}
\mbox{}^{(3)}R(h)=\mbox{}^{(3)}{R}(g)-8\phi^{-1}\Delta_g\phi = 2\Lambda.
\label{EQN_EE}
\end{eqnarray}
If the scaled metric $g_{ij}$ produces a constant curvature $R_g$, the $\phi$ 
are eigenfunctions of the associated Laplace operator,
\begin{eqnarray}
\Delta_g \phi = \frac{1}{8}\left(R_g-2\Lambda\right) \phi.
\end{eqnarray}
We can construct solutions by superposition
\begin{eqnarray}
\phi = \Sigma \mu_i\phi_i,
\label{EQN_PHI}
\end{eqnarray}
$\mu_i>0$, $\Sigma\mu_i=1$, of different eigenfunctions $\phi_i$ to the same
eigenvalue, e.g., those translated in any one of the homogeneous directions. 
The Brill-Lindquist case corresponds to the flat metric $g_{ij}=\delta_{ij}$ 
with $R_g=0$ and three homogeneous directions, whereas the Misner case corresponds 
to the donut metric of (\ref{EQN_BM}) with $R_g=2$ and one homogeneous direction, 
both with $\Lambda=0$. In case of the former, we have the coordinate density
\begin{eqnarray}
\phi_p^2 = \frac{r}{\rho},~~\rho=\sqrt{(x_1-p_1)^2+(x_2-p_2)^2+(x_3-p_3)^2},
\end{eqnarray}
associated with black holes at $p$ (and $r=0$ in the corresponding Schwarzschild
coordinates), where the singular limit $\Lambda\rightarrow0$ recovers
$\phi=1+\frac{m}{2\rho}$. $\Box$.

The eigenvalue problem (\ref{EQN_PHI}) reduces to a Helmholtz
equation when $g_{ij}$ is flat. It then suggests an association to a boson field 
with frequency $\omega = \frac{1}{2}\sqrt{\Lambda}$ when $\Lambda\ge0$.
However, this association does not reflect scale invariance in the 
Riemann tensor, in that it depends only on the log of $\phi$. 

A cosmological event horizon at finite distance alters the spectrum of eigenfunctions.
Whereas the eigenfunctions $r^{-(l+1)}P_l(x)$ to $\Delta \phi=0$ give convergence and 
divergence at asymptotic infinity on the two sheet embedding of the Schwarzschild black 
hole in the Brill-Lindquist case $\Lambda=0$, the eigenfunctions of 
$\Delta\phi = -\frac{1}{4}\Lambda \phi$ for $\Lambda>0$ are periodic across the black hole 
and cosmological event horizons in Figs. \ref{FIG_2a}-\ref{FIG_2b} and are hereby necessarily 
bounded and regular everywhere.

A cosmological event horizon also alters the Hilbert space of radiation states at large distances
from a black hole, and hence the details of Hawking radiation \citep{kan05,zen08}. This is particularly 
pertinent during inflation, when $\Lambda$ is large. The evaporation time of black holes should
remain finite, however, whereby the strong form of the cosmic censorship conjecture described in 
the introduction should continue to hold whenever $m^2\Lambda<\frac{1}{9}$.

A construction similar to (\ref{EQN_L1}-\ref{EQN_L2}) may be pursued for the Misner two-sheet 
embedding of two black holes, based on eigenfunctions $\Delta_D\phi=\frac{1}{4}(1-a^2\Lambda)\phi$, 
where $D$ refers to the donut line-element in (\ref{EQN_BM}) and $a$ refers to the Misner length scale. 
A detailed consideration, however, falls outside of the present discussion. It may also be generized 
to extra dimensions on the basis of the Schwarzschild-de Sitter solution \citep{tan63,gao04}. In 4+1, 
for example, the transformation (\ref{EQN_L1}-\ref{EQN_L2}) now produces a trigonometric expression.

\section{Extended black hole-de Sitter cosmologies}

\begin{figure}
\centerline{\includegraphics{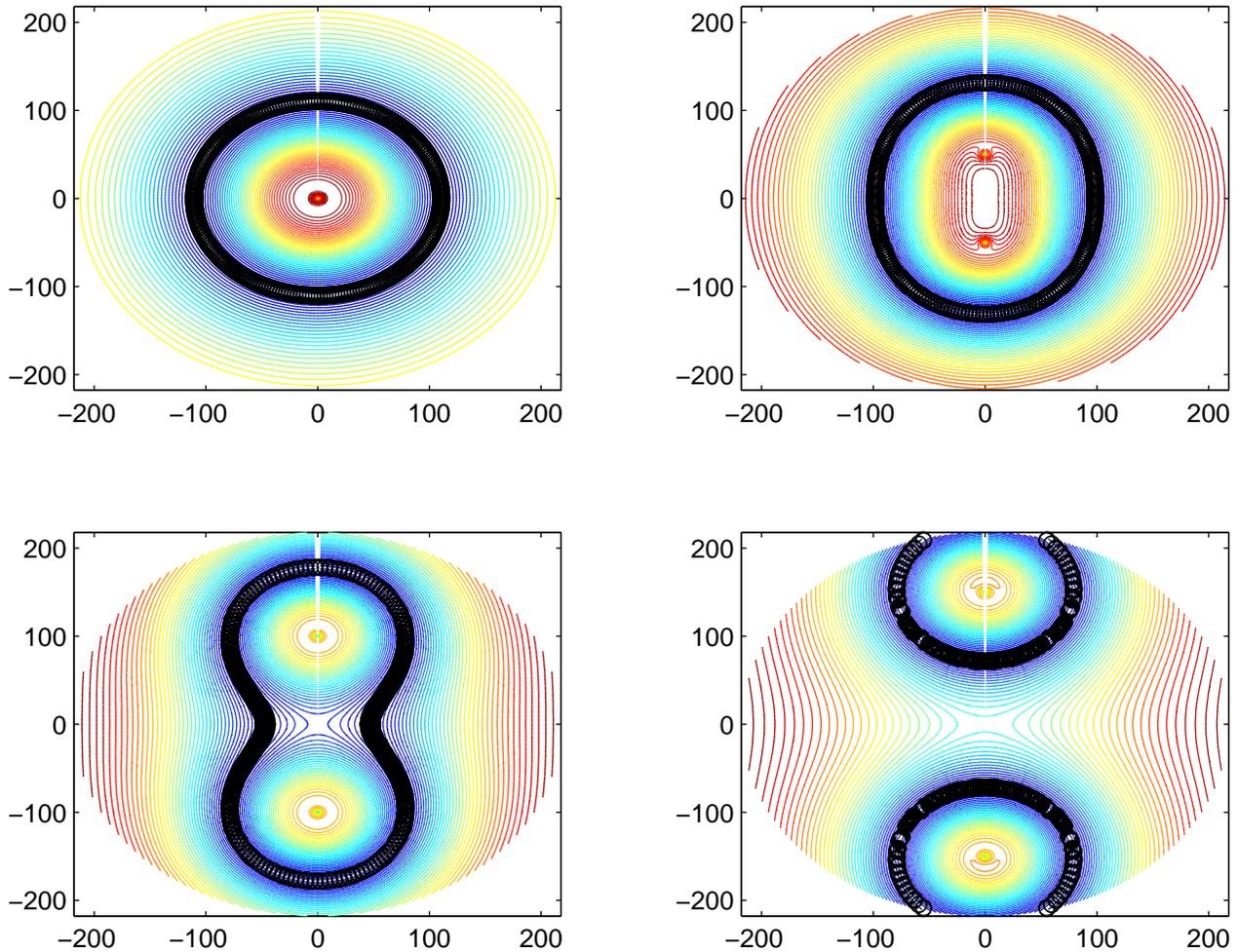}}
\caption{{\small
Initial data of binaries constructed by the generalized superposition principle
starting from zero separation ({\em left top}). Shown are isocurves of constant
$J=|\partial\phi/\partial_\lambda+1/2|$. For small separations, the result is
a black hole binary inside a slightly deformed cosmological event horizon $J=0$
(thick dark line). As the separation increases and becomes large, a binary of two 
universes forms, wherein the second lives on a sheet adjacent to the first and beyond 
its cosmological event horizon corresonding to a dashed sheet in Fig. (\ref{FIG_2b})
({\em right below}). There is an infinity of black hole and cosmological event 
horizons at exponentially increasing coordinate distances and within the two black 
holes (not shown).}}
\label{FIG_4}
\end{figure}

For $0<m^2\Lambda<\frac{1}{9}$, we construct a few binary black hole spacetimes
with different separations as shown in Fig. (\ref{FIG_4}). For small separations, the 
results are very similar to a Brill-Lindquist spacetime inside the cosmological event 
horizon surrounding the binary. For large separations, however, the cosmological event 
horizon splits into two, leading to a binary of two universes. Here, the second universe
lives in the ``tower" of the first, on the sheet beyond its cosmological event 
horizon.

The dynamical evolution of an extended black hole-de Sitter universe with multiple black
holes and cosmological event horizons offers a new route to gravitational radiation from 
the early universe. Here, the waves are generated by the dynamics of primordial black holes, as 
well as by multipole moments of cosmological event horizons. The spectrum of relic waves
may hereby be extended in the infrared, below which what may be expected from mergers of
primordial black holes alone. It poses a novel computational challenge
for numerical relativity, i.e., to compute these relic waves at the present epoch 
from an initial distribution of primordial black hole-de Sitter universes at the
prior to or at the onset of inflation.

\section{Hyperbolicity with curvature-driven lapse}

Spacetime can be described by a foliation in spacelike hypersurfaces, given by
a 3+1 decomposition in terms of $h_{ij}$ \citep{tho86}
\begin{eqnarray}
ds^2 =-N^2dt^2+h_{ij}\left(dx^i+\beta^idt)(dx^j+\beta_jdt\right)
\end{eqnarray}
with lapse function $N$ and shift functions $\beta^i$. The $N$ and $\beta^i$
govern the flow of coordinates in the foliation of hypersurfaces of constant 
coordinate time $t$. In what follows, we shall restrict our attention to normal 
coordinates described by vanishing shift functions, $\beta^i=0$.

In the Hamiltonian equations of motion with vanishing shift function, the
three-metric $h_{ij}$ evolves according to
$\partial_t h_{ij} = - 2NK_{ij} = - NK_i^nh_{nj} - h_{im}NK^m_j$. Integration 
over finite time-intervals $\Delta t$ gives the product
\begin{eqnarray}
h_{ij}(t_i) = \Pi_i e^{-K^{~m}_i N\Delta t}h_{mn}(t_0) e^{-K^n_{~j} N\Delta t},
\label{EQN_H1}
\end{eqnarray}

We propose time-evolution with curvature-driven lapse function, i.e., 
\begin{eqnarray}
D_tN  = -K,~~D_t K_{ij} &=& -D_{ij}N+{\cal K}_{ij}N,
\label{EQN_H2}
\end{eqnarray}
where ${D}_{ij}=D_iD_j-R_{ij}$ and ${\cal K}=KK_{ij} -2K_i^mK_{jm}$.
The gauge condition $D_tN=-K$ is {\em curvature-driven} and is 
distinct from the product of curvature and lapse function in
the harmonic slicing condition $\partial_t N = -N^2 K$ in Eqs. 
(69)-(77) of \cite{abr97}; see also \cite{bro08} for a recent review.

{\bf Theorem.} {\em The 3+1 Hamiltonian evolution equations are hyperbolic
with curvature-driven lapse function $\partial_tN=-K$}.

{\em Proof.} It suffices to consider the problem of small amplitude wave-motion
about flat spacetime, e.g., the asymptotically flat region with
$h_{ij}=\delta_{ij}$ with $N=\pm 1$ at large distances. Here, we have
\begin{eqnarray}
       \partial_t N = -K,~~
\partial_t^2 h_{ij} = -2R_{ij}+2D_iD_jN,~~
     \partial_t^2 K = \Delta K.
\label{EQN_K}
\end{eqnarray}
We recall that \citep{wal84}
\begin{eqnarray}
R_{ij} = -\frac{1}{2} \Delta \delta h_{ij} 
         +\frac{1}{2} \partial_i\partial^e\bar{\delta h}_{ej}
         +\frac{1}{2}\partial_j\partial^e\bar{\delta h}_{ei}
\end{eqnarray}
where $\bar{h}_{ij} = \delta h_{ij} -\frac{1}{2}\delta_{ij} \delta h$, where 
$\delta h=h^{ij}\delta h_{ij}$ refers to the trace of the metric perturbations.

Small amplitude harmonic perturbations about the flat metric are given by
$\delta h_{ij} \sim \hat{h}_{ij}e^{-i\omega t}e^{ik_ix^i}$. Conservation of 
momentum, $D^iK_{ij} = D_jK$, the lapse condition $-i\omega N=- \hat{K},$ 
$\hat{\delta h}_{ij} = - 2i\omega^{-1}\hat{K}_{ij}$ and 
$k^i\hat{K}_{ij} = k_j\hat{K}$, give rise to
\begin{eqnarray}
\partial_i\partial^e\bar{h}_{ej} \rightarrow 
 k_ik^e\hat{h}_{ej} - \frac{1}{2}k_ik_j\hat{{\delta h}}
= i\omega^{-1} (-2 k_ik^e \hat{K}_{ej}+k_ik_j\hat{K})=-i\omega^{-1}k_ik_j\hat{K}.
\end{eqnarray}
We then have
\begin{eqnarray}
\hat{R}_{ij} - \partial_i\partial_j N 
= \frac{1}{2} k^2 \hat{h}_{ij} - i\omega^{-1}k_ik_j\hat{K} + i\omega^{-1}k_ik_j\hat{K}
= \frac{1}{2} k^2 \hat{h}_{ij},
\end{eqnarray}
whereby $\partial_t^2h_{ij}=-2R_{ij}+2D_iD_jN$ gives rise to the dispersion
relation
\begin{eqnarray}
\omega^2=k^2.
\label{EQN_D}
\end{eqnarray}
It follows that all small amplitude metric perturbations propagate along the
light cone, which completes the proof. $\Box$

Clearly, the system (\ref{EQN_K}) is asymptotically stable, as metric
perturbations become small at arbitrarily large distances. Asymptotic 
wave-motion is commonly studied in the so-called transverse traceless 
gauge, or harmonic coordinates--neither of these two coordinate 
conditions are used here.

Numerical integration of (\ref{EQN_H1}-\ref{EQN_H2})
on the donut $(\lambda,\theta,\varphi)$ can be pursued using a conformal decomposition 
to bring out invariance of the aforementioned $C_{ij}$ with respect to scaling in 
$\phi$. To this end, we may set $\phi=e^\eta$, whereby
\begin{eqnarray}
C_{ij} = -2[D_iD_j\eta + g_{ij} \Delta\eta] 
        + 4[D_i\eta D_j\eta - g_{ij} D^p\eta D_p\eta ].
\end{eqnarray}
It follows that $\partial_t K_{ij} = F_{ij}$ with $F_{ij} = 
- N(2K_i^mK_{jm}-KK_{ij}) +  N(R_{ij}+C_{ij}) - D_iD_jN + \Omega_{ij}^k\partial_kN$,
where $\Omega_{ij}^k = 2\phi^{-1}g^{ke}(g_{ie}\partial_j\phi+g_{je}\partial_i\phi-g_{ij}\partial_e\phi)=
2g^{ke}(g_{ie}\partial_j\eta + g_{je}\partial_i\eta - g_{ij}\partial_e\eta)$.

A closed system is obtained by choosing an equation for the conformal
factor. A common choice is $\phi=h^{1/12}$ in terms of the determinant
$h$ of the three metric $h_{ij}$. For outgoing radiation, note that $\phi=1$ 
up to including first order in the wave-amplitude, whereby $\phi=1$ tracks the
merger phase of black hole coalescence (here with overlapping horizon surfaces)
and ringdown in collapse to a single black hole. A slight variation
is to insist $g^{ij}\partial_tg_{ij}=0$. The complete hyperbolic system for 
numerical integration using normal coordinates ($\beta_i=0$) hereby becomes
\begin{eqnarray}
\left\{
\begin{array}{rl}
\partial_t N & = - K,\\
\partial_t \eta & = -\frac{1}{6}NK,\\
\partial_t g_{ij} & = -2N\left(K_{ij} - \frac{1}{3}g_{ij}\tilde{K}_{ij}\right),\\
\partial_t K_{ij} & = F_{ij}(\eta,g_{ij},N),
\end{array}\right.
\end{eqnarray}
where $\tilde{K}_{ij}=e^{-4\eta}K_{ij}$.

The cosmological event horizon $\lambda=\lambda_2$ represents a
{\em turning point} in (\ref{EQN_E1}), representing an extremum
of the radius $\rho\phi^2$ defined by the Neumann condition
\begin{eqnarray}
\frac{d\eta}{d\lambda} = -\frac{1}{2}.
\label{EQN_H22}
\end{eqnarray}
As such, (\ref{EQN_H22}) defines the cosmological event horizon 
$(\lambda_2(\theta,\varphi),\theta,\varphi)$ following superpositions 
(\ref{EQN_PHI}). In dynamical evolutions, it can be used to 
define apparent horizon surfaces. A proper condition of lapse 
function $N$ is that it preserves a simple zero across, i.e., 
the initial condition
\begin{eqnarray}
N=\frac{d\log(\rho\phi^2)}{d\lambda}.
\end{eqnarray}

\section{Conclusions}

We have developed an extension of the black hole-de Sitter space and
extended the superposition principle for the Hamiltonian energy constraint
to (anti-)de Sitter cosmologies with cosmological constant $\Lambda$ in terms of
an eigenvalue problem for the Laplace operator on a metric with constant curvature.

A positive cosmological constant $\Lambda$ is of increasing relevance to our study of
cosmological spacetimes in view of its role in the early universe in view of 
observational evidence for a small positive value at the present epoch.

For one black hole, the conformal scale $\phi$ is periodic, representing 
the extension of the black hole-de Sitter space to an infinite tower of universes. 
The result may become periodic, upon identifying any pair of the these event horizons, 
e.g., $H_0$ with $H_1$ in Fig. 3.
Superposition of two eigenfunctions produces a binary black hole-de Sitter universe with 
one (adjacent) cosmological event horizon when the separation is small relative to the 
cosmological scale $1/\sqrt{\Lambda}$. However, it produces a binary of universes when 
their separation is large described by two adjacent cosmological event horizons.

The results demonstrate extended topologies of black hole-de Sitter cosmologies, wherein
the metric is everywhere smooth, as the physical singularities inherent to black 
holes are moved away into the complex plane and space is given a finite physical
extent. By virtue of a finite evaporation time of black holes by Hawking radiation, there
is a strong cosmic censorship conjecture by which singularities are protected from
direct observation, regardless whether they are located in real or complex coordinate
space.

The dynamical interaction of a binary of two universes (\ref{FIG_4}) is conceivably of 
interest to the early universe and the ensuing generation of primordial gravitational 
waves from an initial distribution of primordial black hole-de Sitter universes, when 
$\Lambda$ was large at the onset of inflation. An extension of our approach to extra 
dimensions is readily given, which is conceivably of interest to dynamical evolution 
of our four-dimensional spacetime within a spacetime of extra dimensions. For computational 
purposes, we give a new hyperbolic formulation of the equations of motion based on a 
curvature-driven lapse function. 

{\bf Acknowledgment.} The author gratefully acknowledges stimulating discussions with
A. Spallicci, M. Volkov, G. Barles, members of the F\'ed\'eration Denis Poisson, and
AEI of the Max Planck Institute, where some of the work was performed. This work is 
supported, in part, by Le Studium IAS of the Universit\'e d'Orl\'eans. The author thanks
the referees for their constructive comments.


\end{document}